\documentclass[11pt,twoside]{article}
\pdfoutput=1
\usepackage{asp2010}
\usepackage{epsf}
\usepackage{epstopdf}
\resetcounters

\begin{document}

\title{Iron and nickel diffusion in subdwarf B stars}
\author{Haili Hu$^1$
\affil{$^1$Institute of Astronomy, The Observatories, Madingley Road, \\Cambridge CB3 0HA}
}

\begin{abstract}
Pulsations in subdwarf B stars are attributed to radiative levitation of iron-group elements in the stellar envelope. Until now, only iron diffusion is accounted for in stellar models used for sdB seismology. However, nickel has also been suggested as a contributor to the opacity bump that drives the pulsation modes. Stellar models including time-dependent atomic diffusion, as we compute here, are needed to evaluate the importance of different iron-group elements for mode driving. 
We perform detailed calculations of radiative accelerations of H, He, C, N, O, Ne, Mg, Fe and Ni and include these in Burgers' diffusion equations. We compute the evolution and non-adiabatic pulsations of a typical subdwarf B star. 
We show that, despite its lower initial abundance, nickel accumulates to comparable mass fractions as iron in the sdB envelope. For accurate determination of pulsation frequencies and mode instability, it is essential that diffusion of both metals are included in stellar models.  The role of other iron-group elements remain to be evaluated.
\end{abstract}

\section{Introduction}
Subdwarf B (sdB) stars have been identified as core He burning stars with an unusual thin H-envelope ($<0.02$ M$_{\odot}$), also called Extreme Horizontal Branch stars because of their location in the Hertzsprung-Russel diagram \citep{heber1986}.  Two subgroups of sdB stars are pulsating with variable class names V361 Hya and V1093 Her. The discovery of short-period (100--400\,s) pulsations in V361 Hya stars by \citet{kilkenny1997} coincided with the prediction of unstable pressure($p$)-mode pulsations in these stars by \citet{charpinet1996}. The excitation was attributed to the opacity mechanism enabled by Fe accumulating diffusively in the stellar envelope \citep{charpinet1997}.
A few years later,  \citet{green2003} discovered long-period pulsations (30-120 min) in the cooler V1093 Her stars. \citet{fontaine2003} showed that the same opacity mechanism can also excite long-period gravity($g$)-modes. However, these authors found a too cool theoretical blue-edge of the $g$-mode's instability strip  compared to observations. The problem can be partly solved by using OP opacities \citep{badnell2005} instead of OPAL's \citep{iglesias1996} \`and by \emph{assuming} that Ni accumulates as well as Fe  \citep{jeffery2006}. Furthermore, \citet{hu2009} showed that by including H-He diffusion the theoretical blue-edge is shifted even closer to the observed value of $30$\,kK. Still, all aforementioned models fail to produce unstable $g$-modes at low spherical degree ($\ell<3$) at the observed temperatures, while these modes are the most likely to be observed owing to geometric cancellation of higher degree modes. It is clear that stellar models with time-dependent diffusion, such as we produce here, are needed in sdB seismology.

\section{Method}
We use a version of the  stellar evolution code \textsc{stars} \citep{eggleton1971} that is adapted for asteroseismology \citep{hu2008}. The code uses nuclear reaction rates from \citet{angulo1999}, except for the $^{14}{\rm N}({\rm p}, \gamma)^{15}{\rm O}$ rate for which the recommended value by \citet{herwig2006} and \citet{formicola2002} is used. Neutrino loss rates are according to \citet{itoh1989,itoh1992}. Convection is treated with a standard mixing-length prescription \citep{bohm-vitense1958} with a mixing length to pressure scale height ratio of $l/H_p=2.0$. Convective mixing is treated as a diffusive process in the framework of mixing length theory. The Rosseland mean opacity $\kappa_{\rm{R}}$ and radiative accelerations $g_{\rm rad}$ are computed in-line during the evolution to account for composition changes consistently. For this we use atomic data and codes from the Opacity Project (OP, \citealt{seaton1994} and \citealt{badnell2005}). For high temperatures outside the OP range ($T>10^8$ K), we use OPAL tables   \citep{iglesias1996} combined with conductive opacities \citep{cassisi2007}.

The starting model for each of our calculations is a ZAEHB model with total mass $M_*=0.46$, envelope mass $M_{\rm env}=3.5\times 10^{-4}$\,M$_{\odot}$ and metallicity $Z=0.02$ with \cite{grevesse1993} metal mixture. We follow the stellar evolution from the start to the end of of core He burning. We compute two series of simulations with atomic diffusion of H, He, C, N, O, Ne, Mg and
\begin{itemize}
\item[(a)] Fe;
\item[(b)] Fe and Ni.
\end{itemize}
The diffusion velocities are computed by solving the \citet{burgers1969} equations with a version of \citet{thoul1994} routine that is adapted to use \citet{paquette1986} resistance coefficients, mean ionic charges and radiative accelerations. 

We use the linear non-adiabatic pulsation code \textsc{mad} \citep{dupret2001} to compute the seismic properties, in particular pulsation frequencies and damping rates, for the stellar evolutionary models. We restrict ourselves to modes of spherical degree $\ell\leq2$, because higher  $\ell$ values are geometrically disfavoured for observation. We search for eigenfrequencies in the range of $0.1\leq\omega\leq20$ where $\omega$ is the dimensionless angular eigenfrequency, $\omega=2\pi f \tau_{\rm{dyn}}$, $f$ is the frequency and $\tau_{\rm dyn}$ is the dynamical time-scale.

The method summarized here is described in more detail by \cite{hu2011}.

\section{Results}
We show the effects of atomic diffusion on the stellar evolutionary track in Sect.~\ref{evolution}. In Sect.~\ref{abundev} the changes in abundances and opacities as result of diffusion are discussed. The effects on mode driving are studied in Sect.~\ref{stability}.

\subsection{Evolutionary tracks}\label{evolution}
\begin{figure}
   \begin{center}
  \includegraphics[width=8cm, angle=-90]{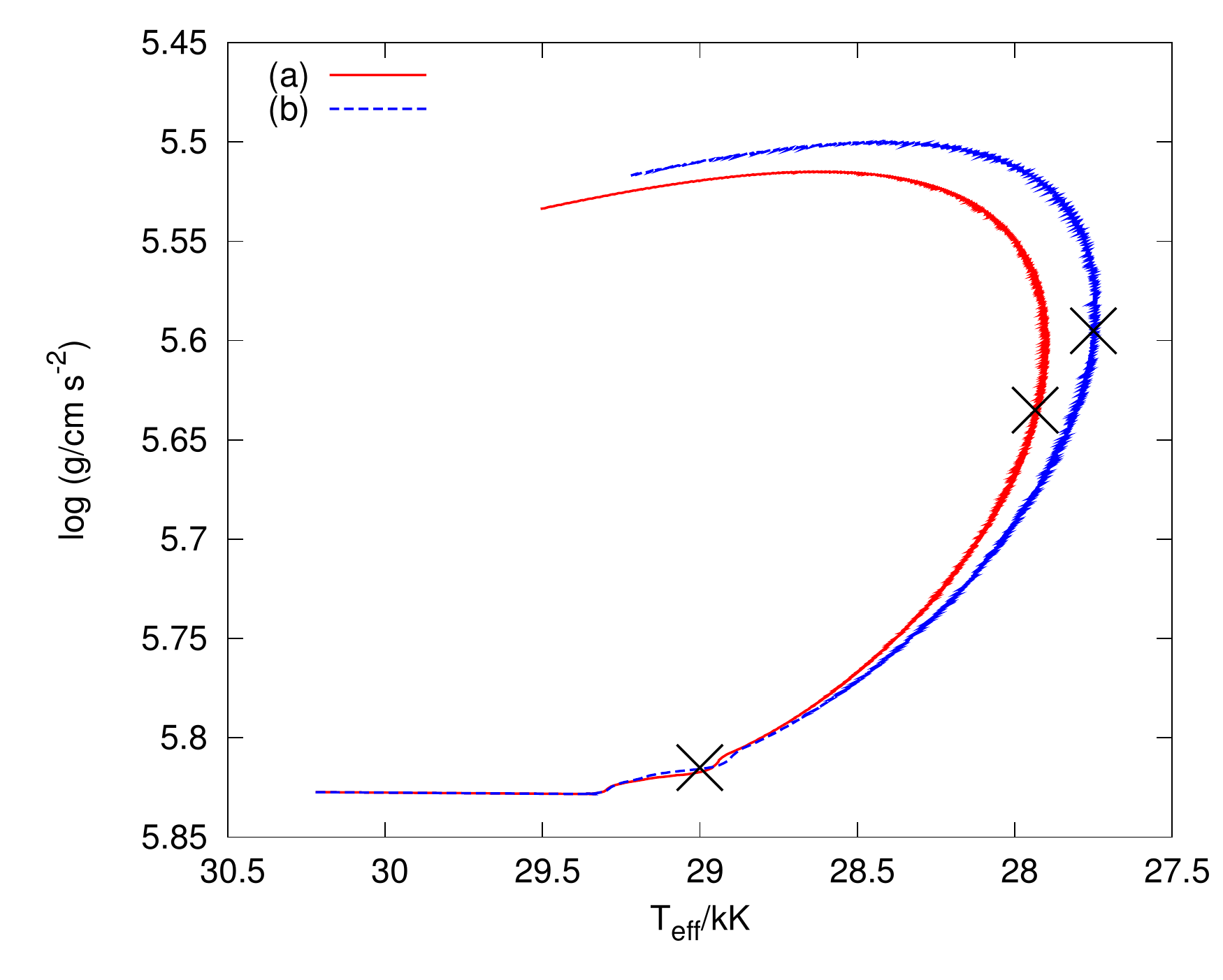}
   \caption{Evolutionary tracks in the $T_{\rm eff}-\log g$ plane for sdB stars with atomic diffusion. Simulation (a) is with diffusion of H, He, C, N, O, Ne, Mg and Fe; simulation (b) is with diffusion of H, He, C, N, O, Ne, Mg, Fe and Ni. The crosses mark the part of the evolution for which the pulsations are shown in Fig.~\ref{mode}.}
              \label{HR}
              \end{center}
\end{figure}

In Fig.~\ref{HR} we plot the evolutionary tracks of the two simulations in a $T_{\rm eff}-\log g$ diagram. We see that including Ni diffusion causes a shift in the track compared to only Fe diffusion. The star becomes slightly cooler and more extended due to an increase in opacity (see Fig.~\ref{opacity}). 
The pulsation frequencies, in particular those of $p$-modes, are very sensitive to $\log g$ and it is therefore important to model diffusion accurately. 

%---------------------------------------
%---------------------------------------

\subsection{Abundances and opacities}\label{abundev}
\begin{figure}
   \begin{center}
  \includegraphics[width=5.75cm, angle=-90]{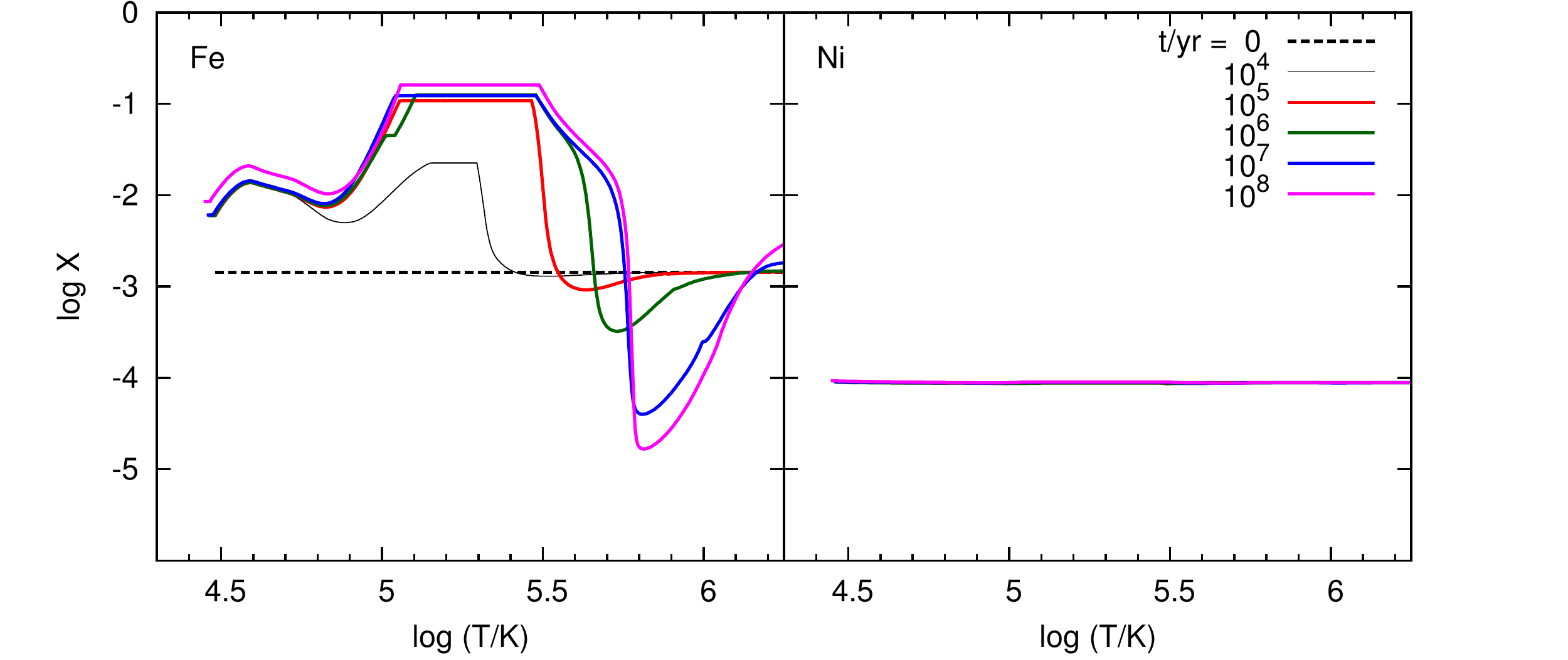}
   \caption{Interior abundance profiles of Fe and Ni at different EHB ages for simulation (a).}
              \label{abunda}
              \end{center}
\end{figure}

\begin{figure}
   \begin{center}
  \includegraphics[width=5.75cm, angle=-90]{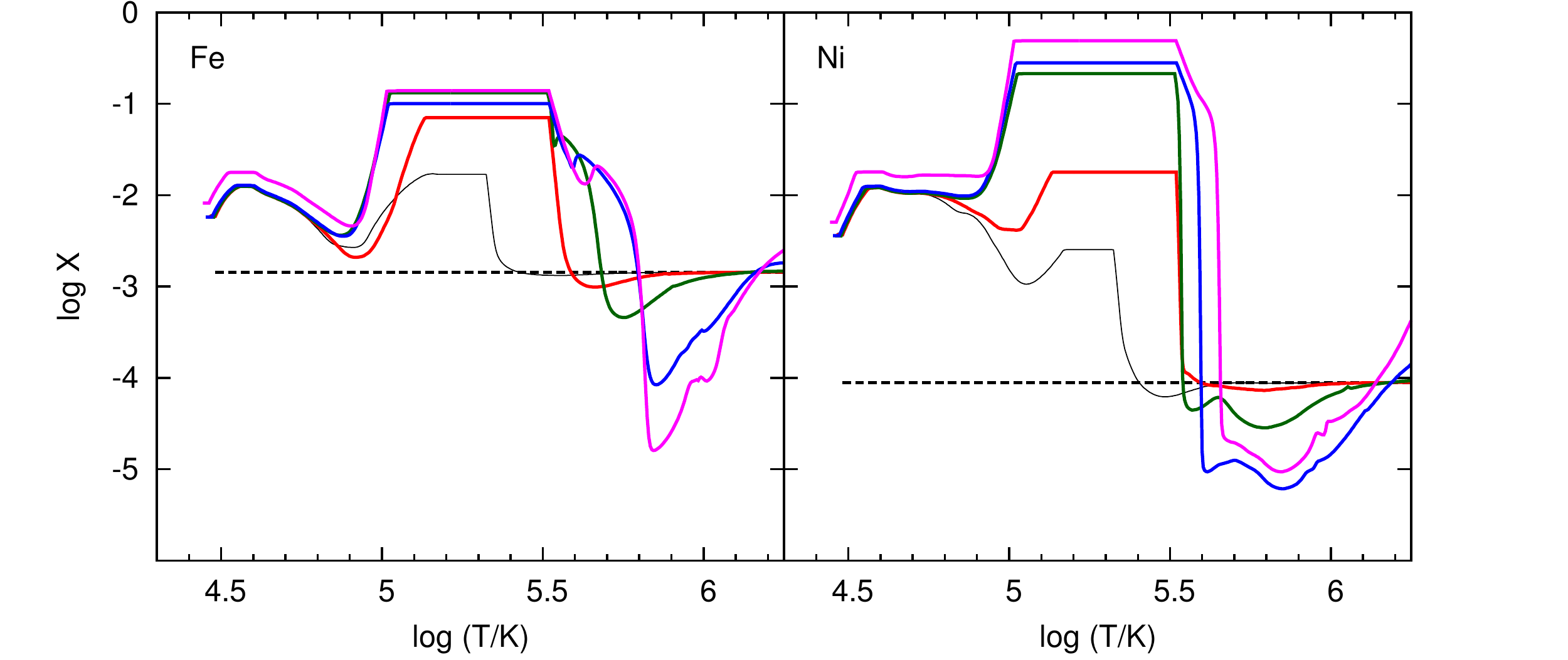}
   \caption{Interior abundance profiles of Fe and Ni at different EHB ages for simulation (b).  Lines are the same as in Fig.~\ref{abunda}.}
              \label{abund}
              \end{center}
\end{figure}

In Fig.~\ref{abunda} and Fig.~\ref{abund}  the time evolution of the interior abundances of Fe and Ni can be followed for simulation (a) and (b), respectively. We see that, despite its lower initial abundance, Ni accumulates to comparable mass fractions as Fe in the sdB envelope. Further, whether or not Ni is allowed to diffuse has an effect on the way Fe accumulates. This is because the chemical composition of the mixture influences the radiative acceleration of each element. After all, the radiative acceleration depends on the Rosseland mean opacity and the monochromatic opacity of the mixture (see e.g.~\citealt{seaton1997}).

\begin{figure}
   \begin{center}
  \includegraphics[width=5.75cm, angle=-90]{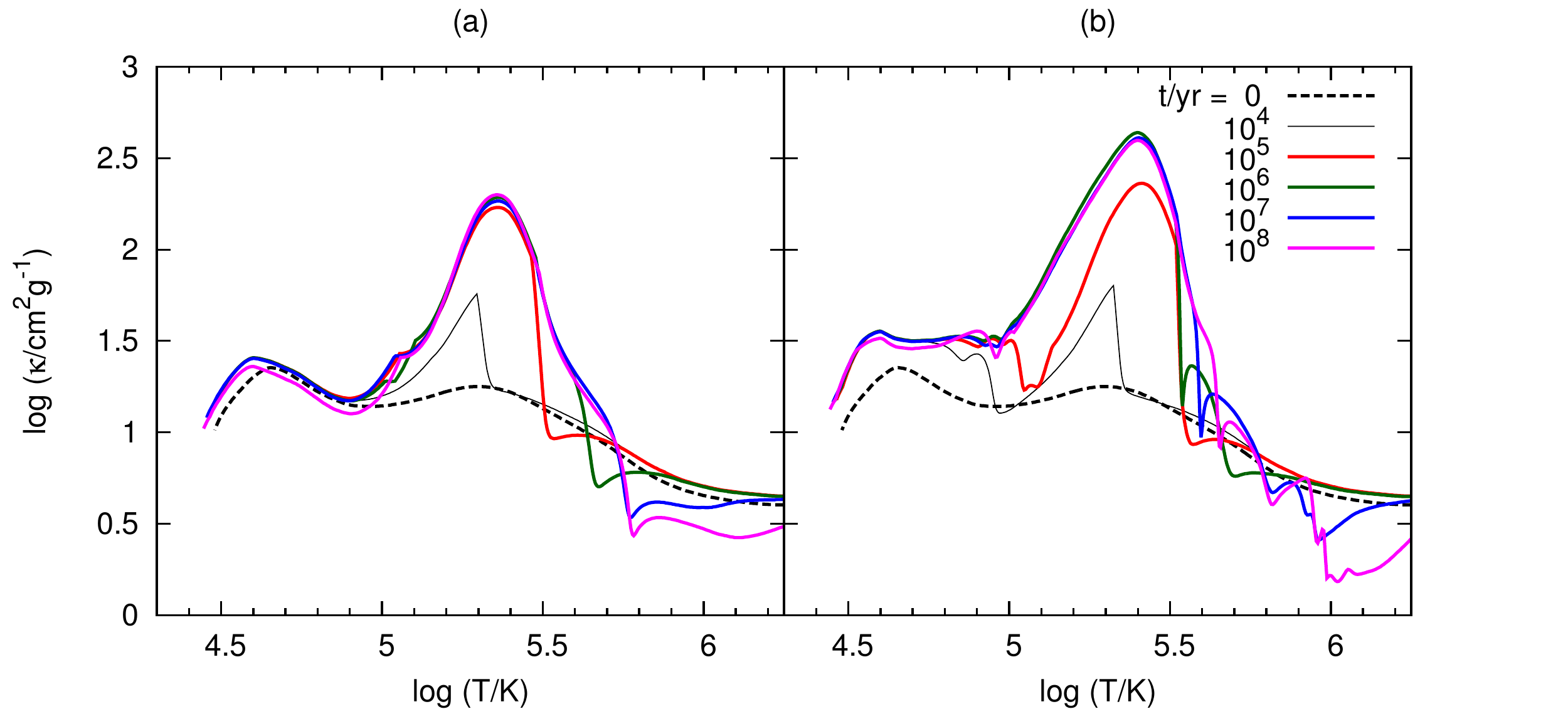}
   \caption{Rosseland mean opacity profiles as a function of temperature at different EHB ages.}
              \label{opacity}
              \end{center}
\end{figure}

The Rosseland mean opacity profiles are shown in Fig.~\ref{opacity}. It can be seen that Ni accumulation causes the opacity bump to occur at slightly higher temperatures as already pointed out by \citet{jeffery2006}. 

%---------------------------------------
%---------------------------------------
%---------------------------------------
%---------------------------------------

%---------------------------------------
%---------------------------------------

\subsection{Stability analysis}\label{stability}
\begin{figure}
   \begin{center}
  \includegraphics[width=5.75cm, angle=-90]{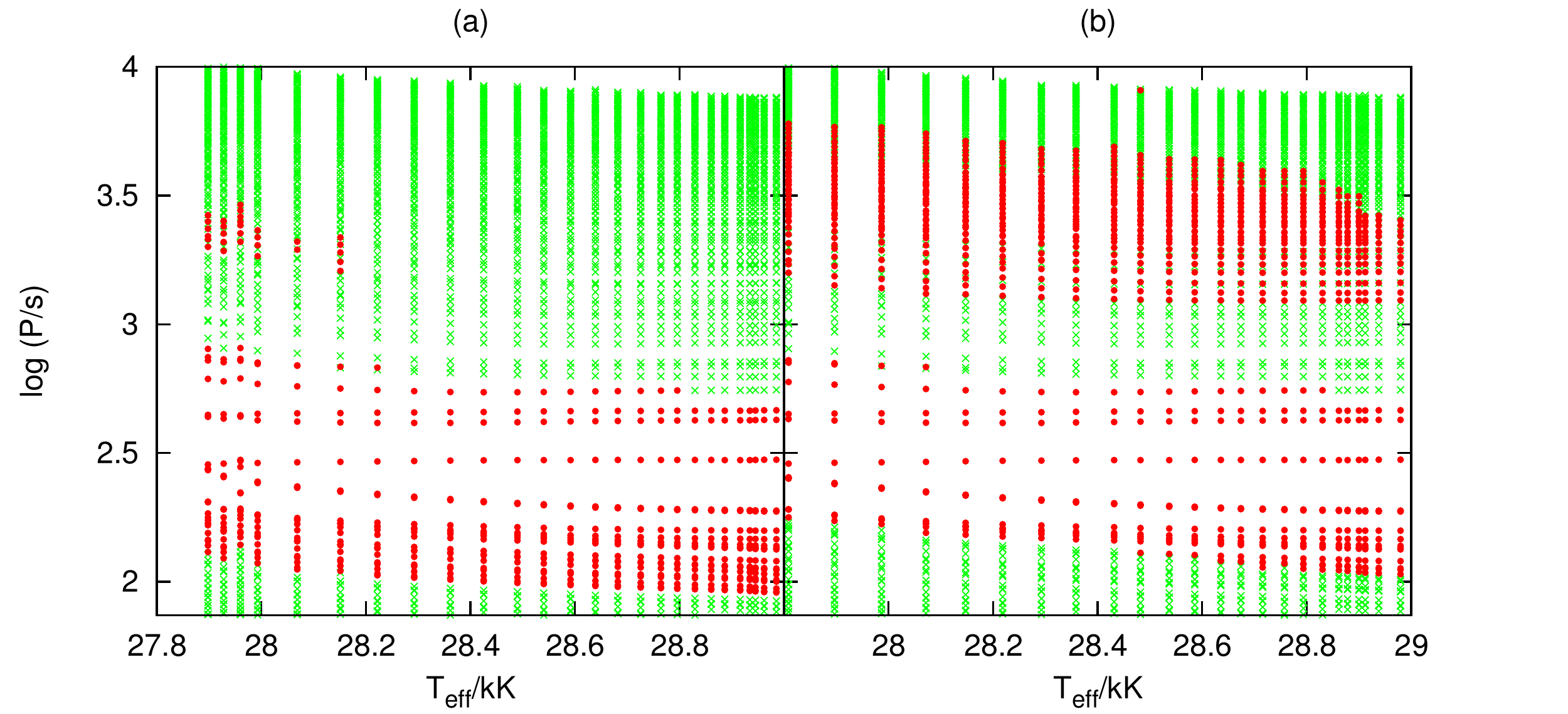}
   \caption{Pulsation periods as a function of effective temperature. Red dots are unstable modes and green crosses are stable. The \emph{left} panel is for simulation (a) and the \emph{right} panel for simulation (b).}
              \label{mode}
              \end{center}
\end{figure}

In Fig.~\ref{mode} we show the pulsation periods as a function of $T_{\rm eff}$. We plot pulsations for only a part of the evolution (indicated in Fig.~\ref{HR}) so that the graphs do not become overcluttered. 
We see that Fe diffusion alone cannot efficiently drive $g$-modes. By including Ni diffusion, many $g$-modes become unstable but the driving of $p$-modes becomes slightly less efficient. This is related to the location of the iron-group opacity bump as shown in Fig.~\ref{opacity} and discussed by \citet{jeffery2006}. It is worth emphasising that we are plotting modes of degree $\ell<3$. These are the first sdB models that show unstable $g$-modes at such high effective temperatures ($T_{\rm eff}\approx29\,{\rm kK}$). In fact, the blue-edge problem of the  V1093\,Her instability strip can finally be solved with our models (Hu et al.~in prep.). 

%---------------------------------------
%---------------------------------------

\section{Conclusions}
We have shown the importance of Ni diffusion, in addition to Fe diffusion, for the study of sdB pulsations. Not only is there a large effect on the mode driving, the stellar evolutionary track is also affected by Ni diffusion. This is expressed in shifts in $T_{\rm eff}$ and  $\log g$ and hence the pulsation frequencies. An accurate prediction of the theoretical pulsation frequencies is essential now that space data from CoRot and Kepler provide high-precision observed frequencies. It may be necessary to treat diffusion of other (iron-group) elements as well, in particular Cr and Mn, and we shall explore this in a forthcoming paper.

\acknowledgements 
HH thanks the conference organization for offering financial support to attend the conference, although she was unable to do so due to the tragic events. HH is supported by the Netherlands Organisation for Scientific Research (NWO).

\bibliographystyle{asp2010}
\bibliography{asphu}

\end{document}